\documentclass{sig-alternate}

\usepackage{microtype}
\usepackage{amsmath}

\usepackage{algorithm}
\usepackage{algorithmic}
\usepackage{bm}

\usepackage{graphicx} 
\usepackage{epsfig}
\usepackage{epstopdf}

\newcommand{\Osymbol}{{\mathcal O}}
\newcommand{\BO}[1]{\Osymbol\left(#1\right)}

\renewcommand{\Pr}[1]{\textrm{\bf Pr}\left[#1\right]}

\newcommand{\footnoteref}[1]{\textsuperscript{\ref{#1}}}

\newtheorem{definition}{Definition}

\begin{document}

\title{Hybrid LSH: Faster Near Neighbors Reporting in High-dimensional Space
\titlenote{Research supported by the Innovation Fund Denmark through the DABAI project}
}

\numberofauthors{1} 
\author{
\alignauthor
Ninh Pham\\
			 \affaddr{Department of Computer Science}\\
       \affaddr{University of Copenhagen}\\
       \affaddr{Denmark}\\
       \email{pham@di.ku.dk}
}

\maketitle

\begin{abstract}

We study the $r$-near neighbors reporting problem ($r$NNR) (or \textit{spherical range reporting}), i.e., reporting \emph{all} points in a high-dimensional point set $S$ that lie within a radius $r$ of a given query point. This problem has played building block roles in finding near-duplicate web pages, solving $k$-diverse near neighbor search and content-based image retrieval problems. Our approach builds upon the locality-sensitive hashing (LSH) framework due to its appealing asymptotic sub-linear query time for near neighbor search problems in high-dimensional space. A bottleneck of the traditional LSH scheme for solving $r$NNR is that its performance is sensitive to data and query-dependent parameters. On data sets whose data distributions have diverse local density patterns, LSH with inappropriate tuning parameters can sometimes be outperformed by a simple linear search. 

In this paper, we introduce a hybrid search strategy between LSH-based search and linear search for $r$NNR in high-dimensional space. By integrating an auxiliary data structure into LSH hash tables, we can efficiently estimate the computational cost of LSH-based search for a given query regardless of the data distribution. This means that we are able to choose the appropriate search strategy between LSH-based search and linear search to achieve better performance. Moreover, the integrated data structure is time efficient and fits well with many recent state-of-the-art LSH-based approaches. Our experiments on real-world data sets show that the hybrid search approach outperforms (or is comparable to) both LSH-based search and linear search for a wide range of search radii and data distributions in high-dimensional space.

\end{abstract}




\section{Introduction}

We study the $r$-near neighbors reporting problem ($r$NNR) (or \textit{spherical range reporting})~\cite{ArXiv, Arya_ESA10}: \textit{Given a $d$-dimensional point set $S$ of size $n$, reporting \emph{all} points in $S$ that lie within a radius $r$ of a given query point}. This problem has played building block roles in finding near-duplicate web pages~\cite{Henzinger_SIGIR06}, solving $k$-diverse near neighbor search~\cite{Abbar_WWW13} and content-based image retrieval problems~\cite{Yu_ICML14}. Recent theoretical work~\cite{ArXiv, Alman_FOCS15} conjectures that solving $r$NNR exactly in time truly sub-linear in $n$ seems to demand space exponential in $d$, which is an example of the phenomenon ``curse of dimensionality''.

Since exact solutions of $r$NNR generally degrade as dimensionality increases, we investigate an \emph{approximate} variant of $r$NNR. That is, given a parameter $0 < \delta < 1$, we allow the algorithm to return each point in $S$ that lie within a radius $r$ of the query point  with probability $1 - \delta$. Our approach builds upon on the \emph{locality-sensitive hashing} (LSH)~\cite{Andoni_CACM08, Indyk_STOC98}, one of the most widely used solution for near neighbor search problems. In a nutshell, LSH hashes near points into the same bucket with good probability, and increases the gap of collision probability between near and far points. It typically needs to use multiple hash tables to obtain probabilistic guarantees. Search candidates are \emph{distinct} data points that are hashed into the same bucket as the query in hash tables.

\begin{figure} [t]
\centering
\includegraphics[width=0.7\columnwidth]{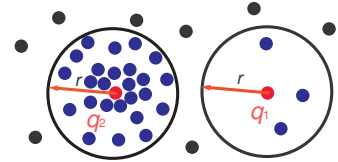}
\caption{An example of LSH bottleneck. Given a radius $r$, LSH works efficiently with the query $\bm q_1$ on sparse area, since it will report just a few points. However, LSH is worse than linear search with the ``hard'' query $\bm q_2$ on dense area. Since the output size of $\bm q_2$ is nearly the data set size and many points are very close to $\bm q_2$, duplicates show up in most hash tables and the cost of removing duplicates will be the computational bottleneck.}
\label{fig:example}
\end{figure}

Since its first introduction, several LSH schemes~\cite{Broder_STOC98, Charikar_STOC02, Datar_SOCG04, Gionis_VLDB99, Lv_VLDB07, Pagh_SODA16} have been proposed for a wide range of metric distances in high-dimensional space. However, a bottleneck of using LSH for solving $r$NNR is that its performance is sensitive to the parameters which depend on the distance distribution between data points and query points. Such parameters are hard to tune on data sets whose data distributions have diverse \emph{local} density patterns. Figure~\ref{fig:example} shows an illustration of this bottleneck. 

In practice, LSH needs to use significant space (i.e., hundreds of hash tables)~\cite{Gionis_VLDB99} or the multi-probe approach~\cite{Lv_VLDB07} which examines several ``close'' buckets in a hash table. In other words, the number of examined buckets needs to be sufficiently large to obtain high accuracy. In turn, the cost of removing duplicates (i.e., points colliding with the query in several hash tables) turns out to be the computational bottleneck when there are many points close to the query. This observation has been shown on the Webspam dataset in the experiment section even with very small radii.
%

In this work, we study a hybrid search strategy between LSH-based search and linear search for $r$NNR in an arbitrary high-dimensional space and distance measure that allows LSH. By integrating the so-called HyperLogLog data structures~\cite{Fusy_AOFA07} into LSH hash tables, we can quickly and accurately estimate the output size and derive the computational cost of LSH-based search for a given query point regardless of the data distribution. In other words, we are able to choose the appropriate search strategy between LSH-based search and linear search to achieve better performance (i.e., running time and recall ratio). Moreover, the proposed solution can be adapted to many recent state-of-the-art LSH-based approaches~\cite{ArXiv, Lv_VLDB07, Pagh_SODA16}. Our experiments on real-world datasets demonstrate that the proposed hybrid search outperforms (or is comparable to) both LSH-based search and linear search for a wide range of search radii and data distributions in high-dimensional space.


\section{Background and Preliminaries}\label{sec:preliminaries}

\textbf{Problem setting.} Our problem, $r$-near neighbor reporting under any distance measure, is defined as follows:
\begin{definition}($r$-near neighbor reporting or $r$NNR)
Given a set $S \subset {\bf R}^d$, $|S| = n$, a distance function $f$, and parameters $r > 0$, $\delta > 0$, construct a data structure that, given any query $\bm{q} \in {\bf R}^d$, return \emph{each} point $\bm{x} \in S$ where $f(\bm{x}, \bm{q}) \leq r$ with probability $1 - \delta$. 
\end{definition}

We call this the ``exact'' $r$NNR problem in case $\delta = 0$, otherwise it is the ``approximate'' variant. 

\textbf{Locality-sensitive hashing (LSH).} LSH can be used for solving approximate $r$NNR in high-dimensional space because its running time is usually better than linear search with appropriate tuning parameters~\cite{Andoni_CACM08}.

%
\begin{definition}\emph{(Indyk and Motwani~\cite{Indyk_STOC98})}
\label{def:LSH}
Fix a distance function $f: {\bf R}^d \times {\bf R}^d \rightarrow {\bf R}$.
For positive reals $r$, $c$, $p_1$, $p_2$, and $p_1 > p_2$, $c > 1$, a family of functions $\mathcal{H}$ is \emph{$(r,cr,p_1,p_2)$-sensitive} if for uniformly chosen $h \in \mathcal{H}$ and all $\bm{x}, \bm{y}\in {\bf R}^d$:
\begin{itemize}
	\item If $f(\bm{x}, \bm{y}) \leq r$ then $\Pr{h(\bm{x})=h(\bm{y})} \geq p_1$;
	\item If $f(\bm{x}, \bm{y}) \geq cr$ then $\Pr{h(\bm{x})=h(\bm{y})} \leq p_2$.
\end{itemize}
\end{definition}

Given an LSH family $\mathcal{H}$, the classic LSH algorithm constructs $L$ hash tables by hashing data points using $L$ hash functions $g_j$, $j = 1, \ldots, L$, by setting $g_j = \left( h_{j}^{1}, \ldots, h_{j}^{k} \right)$, where $h_j^{i}$, $i = 1, \ldots, k$, are chosen randomly from the LSH family $\mathcal{H}$. Concatenating $k$ such random hash functions $h_j^{i}$ increases the gap of collision probability between near points and far points. To process a query $\bm{q}$, one needs to get a candidate set by retrieving all points from the bucket $g_j(\bm{q})$ in the $j$th hash table, $j = 1, \ldots, L$. Each \emph{distinct} point $\bm{x}$ in the candidate set is reported if $f(\bm{x}, \bm{q}) \leq r$.

For the approximate $r$NNR, a near neighbor has to be reported with a probability at least $1 - \delta$. Hence, one can fix the number of hash tables, $L$, and set the value $k$ as a function of $L$ and $\delta$. A simple computation indicates that $k = \left\lceil \log{(1 - \delta^{1/L})} / \log{p_1}\right\rceil$ leads to good performance\footnote{This is a practical setting used in E2LSH package (http://www.mit.edu/$\sim$andoni/LSH/)}. Note that our parameter setting is different from the standard setting $k = \log{n}, L = n^{\rho}$, where $\rho = \log{p_1}/\log{p_2}$~\cite{Indyk_STOC98}, since we focus on reporting \emph{every} $r$-near neighbor.

Although LSH-based algorithm can efficiently solve $r$NNR problem, it might run in $\BO{nL}$ time in the worst case, see Figure~\ref{fig:example} as an example. Tuning appropriate parameters $k, L$ for a given dataset whose data distribution has diverse local density patterns remains a tedious process.

\textbf{HyperLogLog (HLL) for count-distinct problem.} While counting the exact number of distinct elements in a data stream is simple with space linear to the cardinality, approximating such the cardinality using limited memory is an important problem with broad industrial applications. Among efficient algorithms for the problem, HyperLogLog (HLL)~\cite{Fusy_AOFA07} constitutes the state-of-the-art (i.e., a near-optimal probabilistic algorithm) when there is no prior estimate of the cardinality. This means that it achieves a superior accuracy for a given fixed amount of memory over other techniques. 

HLL builds an array $M$ of $m$ zero registers. For an element $i$, it generates a random \emph{integer} pair $\{m_i, v_i\}$ where $m_i \sim \texttt{Uniform}([m])$ indicates a position in $M$, and $v_i \sim \texttt{Geometric}(1/2)$ is an update value. The array $M$ updates the value at the position $m_i$ by $\max{(M[m_i], v_i)}$. After processing all elements, the cardinality estimator of the stream is $\theta_m m^2 \left( \sum_{j=1}^{m}2^{-M[j]}\right)$, where $\theta_m$ is a constant to correct the bias. HLL works optimally with \emph{distributed} data streams since we can merge several HLLs by collecting register values and applying component-wise a \texttt{max} operation. The relative error of HLL is $1.04/\sqrt{m}$. More details of the theoretical analysis and a practical version of HLL can be seen in~\cite{Fusy_AOFA07}. 

\section{Algorithm}\label{sec:algorithm}

This section describes our novel hybrid search strategy which interchanges LSH-based search and linear search for solving $r$NNR. We first present a simple but accurate computational cost model to measure the performance of LSH-based search. By constructing an HLL data structure in each bucket of hash tables, we are able to estimate the computational cost of LSH-based search, and then identify the condition whether LSH-based search or linear search is used.

\subsection{Computational Cost Model}\label{sec:costmodel}

For each query, LSH-based search needs to process following operations: (1) Step S1: Compute hash functions to identify the bucket of query in $L$ hash tables, (2) Step S2: Look up in each hash table the points of the same bucket of query, and merge them together for removing duplicate to form a candidate set, and (3) Step S3: Compute the distance between candidates and the query to report near neighbor points. Typically, the cost of~S1 is very small and dominated by the cost of~S2 and~S3, which significantly depend on the distance distribution between the query and data points. 

To process Step~S2, one typically uses a hash table or a bitvector of $n$ bits to store non-duplicate entries. The cost of such techniques is proportional to the total number of collisions ($\#collisions$) encountered in $L$ hash tables, which can be directly computed by simply storing the bucket size. The cost of~S3 is clearly proportional to the candidate set size ($candSize$). The total cost of LSH-based search is composed of the cost of~S2 and~S3, as formalized in Equation~(\ref{eq:lsh}).

Given $\alpha$ as the average cost of removing a duplicate, and $\beta$ as the cost of a distance computation, we formalize the total cost of LSH-based search and linear search as follows:
\begin{align} 
\textbf{LSHCost} &= \alpha \cdot \#collisions + \beta \cdot candSize \label{eq:lsh} \\
\textbf{LinearCost} &= \beta \cdot n \label{eq:linear}
\end{align}

Given such constants $\alpha, \beta$, we can compute exactly \textbf{LinearCost}, but we need $candSize$ for computing \textbf{LSHCost}. By constructing an HLL data structure for each bucket, we can derive the HLL of the candidate set. Therefore, we can accurately approximate $candSize$, and then estimate the \textbf{LSHCost}. In turn, we can compare \textbf{LinearCost} and \textbf{LSHCost} in order to \emph{interchange} LSH-based search with linear search to achieve better performance.

\subsection{Hybrid Search Strategy}\label{sec:hybrid}

We construct an HLL for each bucket when building LSH hash tables, as shown in Algorithm~\ref{alg:construct}. Given a query $\bm q$, we view point indexes hashed in the buckets $g_1(\bm{q}), \cdots, g_L(\bm{q})$ as $L$ partitions of a data stream. We will estimate the number of distinct elements of such data stream, which is the $candSize$ in Equation~(\ref{eq:lsh}). By estimating \textbf{LSHCost} and comparing it to \textbf{LinearCost}, we can identify the suitable search strategy, as shown in Algorithm~\ref{alg:query}.

\begin{algorithm}[ht]
\caption{Construct LSH hash tables}
\label{alg:construct} 									
\begin{algorithmic} [1]
\REQUIRE {A point set $S$, and $L$ hash functions: $g_1,\ldots, g_L$}
\FOR {each $\bm{x} \in S$}
	\FOR {each hash table $T_i$ using hash function $g_i$}
		\STATE{Insert $\bm{x}$ into the bucket $g_i(\bm{x})$}
		\STATE{Update HyperLogLog of the bucket $g_i(\bm{x})$}		
	\ENDFOR
\ENDFOR
\normalsize
\end{algorithmic}
\end{algorithm}
%
\begin{algorithm}[ht]
\caption{Hybrid search for $r$-NN}
\label{alg:query} 									
\begin{algorithmic} [1]
\REQUIRE {A query point $\bm q$, and $L$ hash tables: $T_1,\ldots, T_L$}
\STATE {Get the size of the buckets $g_1(\bm{q}),\ldots, g_L(\bm{q})$ to compute $\#collisions$}
\STATE {Merge HLLs of the buckets $g_1(\bm{q}),\ldots, g_L(\bm{q})$ to estimate $candSize$}
\STATE {Estimate \textbf{LSHCost} using Equation~(\ref{eq:lsh}), and compute \textbf{LinearCost} using Equation~(\ref{eq:linear})}
\STATE {Choose LSH-based search if \textbf{LSHCost} < \textbf{LinearCost}; otherwise, use linear search}
\normalsize
\end{algorithmic}
\end{algorithm}
%

\textbf{The time complexity analysis.} 
Now, we analyze the complexity of the two algorithms. Algorithm~\ref{alg:construct} uses a space overhead due to the additional HLLs. For each bucket, an HLL needs $O(m)$ space where $m$ is the number of registers of HLL, which governs the accuracy of the $candSize$ estimate. In practice, we only need $m = 32 - 128$. This means that the space overhead of HLLs is usually smaller than large buckets (e.g., $\#points > m$). For small buckets (e.g., $\#points < m$), we might not need HLL, since we can update the merged HLL on demand at the query time. This trick can save the space overhead and improve the running time of the algorithm.

Algorithm~\ref{alg:query} is more important since it governs the running time of the algorithm. Compared to the classic LSH-based search, the additional cost of the hybrid search approach is from merging $L$ HLL data structures and estimating $candSize$, which takes $O(mL)$. Such cost is often smaller than (or comparable to) the cost of Step~S1, i.e., hash functions computation on LSH families~\cite{Broder_STOC98, Charikar_STOC02, Datar_SOCG04, Indyk_STOC98}. 
In other words, the cost overhead caused by our hybrid search approach is little and dominated by the total search cost. 

\section{Experiment}\label{sec:experiment}

We implemented algorithms in Python~3 and conducted experiments on an Intel Xeon Processor E5-1650~v3 with~64GB of RAM.
We compared the performance of different search strategies, including hybrid search, LSH-based search, and linear search for reporting near neighbors on several metric distances allowing LSH. 
We used~4 real-world data sets: Corel Images\footnote{https://archive.ics.uci.edu/ml/datasets/\label{note1}} ($n = 68,040, d =32$), CoverType\footnoteref{note1} ($n = 581,012, d = 54$), Webspam\footnote{http://www.csie.ntu.edu.tw/$\sim$cjlin/libsvmtools/datasets/\label{note2}} ($n = 350,000, d = 254$), and MNIST\footnoteref{note2} ($n = 60,000, d = 780$). 
For each dataset, we randomly remove~100 points and use it as the query set, and report the average of~5 runs of algorithms on the query set.

For each metric distance, we use the corresponding LSH family. Particularly, we applied SimHash~\cite{Charikar_STOC02} to obtain~64-bit fingerprint vectors for MNIST and use bit sampling LSH~\cite{Indyk_STOC98} for Hamming distance. CoverType and Corel Images use random projection-based LSH~\cite{Datar_SOCG04} for~L1 and~L2 distances, respectively. Webspam uses SimHash~\cite{Charikar_STOC02} for cosine distance.

\begin{figure*} [ht]
\centering
\includegraphics[width=1.0\textwidth]{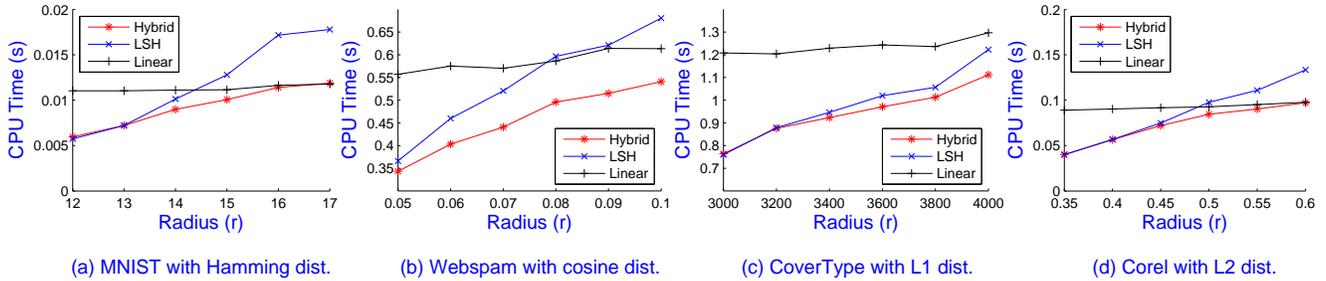}
\caption{Comparison of CPU Time (s) for a query set between hybrid search (Hybrid), LSH-based search (LSH), and linear search (Linear) on~4 data sets using different metric distances.}
\label{fig:lsh}
\end{figure*}

\subsection{Efficiency of HyperLogLog}

This subsection presents experiments to evaluate the efficiency of HLLs on estimating the candidate set size for a given query point. For  HLL's parameter, we fix $m = 128$ to achieve a relative error at most 10\% as suggested in~\cite{Fusy_AOFA07}. For LSH's parameters, we fix $L = 50$ and set $k = \left\lceil \log{(1 - \delta^{1/L})} / \log{p_1}\right\rceil$, where $\delta = 10\%$ and $p_1$ is the collision probability for points within the radius $r$ to the query. This setting is used for SimHash~\cite{Charikar_STOC02} and bit sampling LSH~\cite{Indyk_STOC98}. For random projection-based LSH~\cite{Datar_SOCG04} for L1 and L2 distances, in order to achieve $\delta = 10\%$, we have to adjust $k = 8, w = 4r$ and $k = 7, w = 2r$, respectively, where $w$ is an additional parameter of such LSHs. We note that HLL estimation takes $O(mL)$ time, so this cost is almost constant when fixing $m$ and $L$.

\begin{table}[h]
\centering
\caption{Relative cost and  error of HLLs}
\label{tb:HLL} 
\begin{tabular}{|c|c|c|c|c|c|} \hline
Dataset & Webspam & CoverType & Corel & MNIST \\ \hline
\% Cost & 1.31\% & 0.12\% & 3.18\% &  17.54\% \\ \hline
\% Error & 5.99\% & 5.86\% & 6.74\% & 6.8\% \\ \hline
\end{tabular}
\end{table}

Table~\ref{tb:HLL} shows the average performance of HLL over~4 datasets for a small range of radii where LSH-based search significantly outperforms linear search. It is clear that the cost of HLL is very little, less than 4\% of the total cost for the real-value data points. For MNIST, since the distance computation cost is very cheap due to binary representation, the cost of HLL is 17.54\% of the total cost. However, since MNIST is very small ($n = 60000$), we can set $m = 32$ to reduce the cost to 4.4\% without degrading the performance. 

Regarding the accuracy, although theoretical analysis guarantees a relative error of 10\%, the practical relative error is even much smaller, less than 7\% with standard deviation around 5\% for all datasets. The small overhead cost and high accuracy provided by HLL enables us to efficiently estimate the total cost of LSH-based search, see Equation~(\ref{eq:lsh}), and identify the appropriate search strategy.

\subsection{Efficiency of Hybrid Search}
 
\begin{figure} [ht]
\centering
\includegraphics[width=1.0\columnwidth]{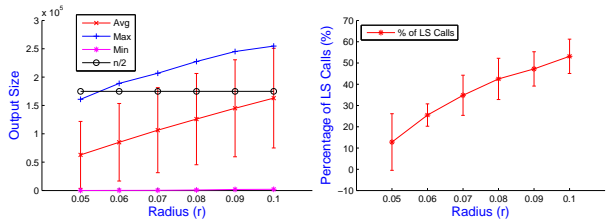}
\caption{Left: Average, maximum, and minimum output size of queries; Right: Percentage of linear search (LS) calls used in hybrid search for Webspam.}
\label{fig:webspam}
\end{figure}

This subsection studies the performance of our proposed hybrid search strategy. 
To compare \textbf{LSHCost} and \textbf{LinearCost}, we need to identify the ratio $\beta / \alpha$, which obviously depends on the implementation, the sparsity of the dataset and the used distance metric. 
We use a random set of 100 queries and 10,000 data points for choosing the ratio $\beta / \alpha$ as~10, 10, 6, 1 for Webspam, Covertype, Corel, and MNIST, respectively. We use the same setting as the previous section for LSH's and HLL's parameters.

Figure~\ref{fig:lsh} shows the average running time in seconds of the~3 search strategies. For small $r$, LSH-based search and hybrid search are comparable, but superior to linear search since the output size of each query is rather small. When $r$ increases, hybrid search gains substantial advantages by interchanging LSH-based search with linear search since there are more ``hard'' queries on the query set. It outperforms LSH-based search and eventually converges to linear search. Specifically, hybrid search provides superior performance compared to both LSH-based search and linear search on Webspam, as shown in Figure~\ref{fig:lsh}.b. This is due to the fact that Webspam has several ``hard'' queries for even very small radii ($r \leq 0.1$).

Figure~\ref{fig:webspam} reveals that the output size varies significantly even with small $r$. 
The maximum output size is almost more than half of the point set size ($n/2$) whereas the minimum output size is very tiny. 
This means that Webspam has many ``hard'' queries, and therefore hybrid search gives superior average performance. 
The right figure confirms this observation by showing the average percentage of linear search calls for hybrid search. 
This amount is at least 10\% at $r=0.05$ and increases to approximate 50\% at $r=0.1$.

We note that hybrid search gives higher recall ratio than LSH-based search since it uses linear search for ``hard'' queries. Due to the limit of space, we do not report it here.

\section{Conclusions}\label{sec:conclusion}

In this paper, we propose a hybrid search strategy for LSH on $r$NNR problem in high-dimensional space. By integrating an HyperLogLog data structure for each bucket, we can estimate the total cost of LSH-based search and choose the appropriate search strategy between LSH-based search and linear search to achieve better performance. Our experiments on real-world data sets demonstrate that the proposed approach outperforms (or is comparable to) both LSH-based search and linear search for a wide range of search radii and data distributions in high-dimensional space. We observed that our hybrid search fits well with the multi-probe LSH schemes~\cite{ArXiv, Lv_VLDB07} and the covering LSH~\cite{Pagh_SODA16}, which typically require a large number of probes. Applying hybrid search on these LSH schemes for $r$NNS will be our future work.



\section{Acknowledgments}
Part of the work was done while the author was working at IT University of Copenhagen through the SSS project.
The author would like to thank Rasmus Pagh and Martin Aum\"{u}ller for useful comments, and members of the SSS project for helpful discussions.

%


\bibliographystyle{abbrv}
\bibliography{sigproc}
%
%

\end{document}